\documentclass[useAMS,usenatbib]{mn2e}
\usepackage{graphicx}
\usepackage{lineno}	

\title[Differential rotation measurement of soft X-Ray corona] {Differential rotation measurement of soft X-Ray corona}

\author[Satish Chandra, Hari Om Vats and K. N. Iyer]{Satish Chandra$^{1}$\thanks{
satish0402@gmail.com} Hari Om Vats$^{2}$\thanks{vats@prl.res.in} and K. N.
Iyer$^{3}$\thanks{iyerkn@yahoo.com}\\
$^{1}$Department of Physics, PPN College, Kanpur - 208 001, INDIA.\\
$^{2}$Physical Research Laboratory, Ahmedabad - 380 009, INDIA.\\
$^{3}$Department of Physics, Saurastra University, Rajkot - 360 005, INDIA.}
\begin{document}

\date{Accepted for publication in MNRAS}

\pagerange{\pageref{firstpage}--\pageref{lastpage}} \pubyear{8888}

\maketitle


\label{firstpage}

\begin{abstract}
The aim of this paper is to study the latitudinal variation in the solar rotation in soft X-ray corona. The time series bins are formed on different latitude regions of the solar full disk (SFD) images that extend from 80\degr S to 80\degr N. These SFD images are obtained with the soft X-ray telescope (SXT) on board the {\it Yohkoh} solar observatory. The autocorrelation analyses are performed with the time series that track the SXR flux modulations in the solar corona. Then for each year, extending from 1992 to 2001, we obtain the coronal sidereal rotation rate as a function of the latitude. The present analysis from SXR radiation reveals that; (i) the equatorial rotation rate of the corona is comparable to the rotation rate of the photosphere and the chromosphere, (ii) the differential profile with respect to the latitude varies throughout the period of the study; it is more in the year 1999 and least in 1994  and (iii) the equatorial rotation period varies systematically with sunspot numbers and indicates its dependence on the phases of the solar activity cycle.
\end{abstract}

\begin{keywords}
Sun: corona -- Sun: X-rays -- Sun: rotation
\end{keywords}

\section{Introduction}

\begin{figure*}
  \centering{
  \includegraphics[width=\textwidth]{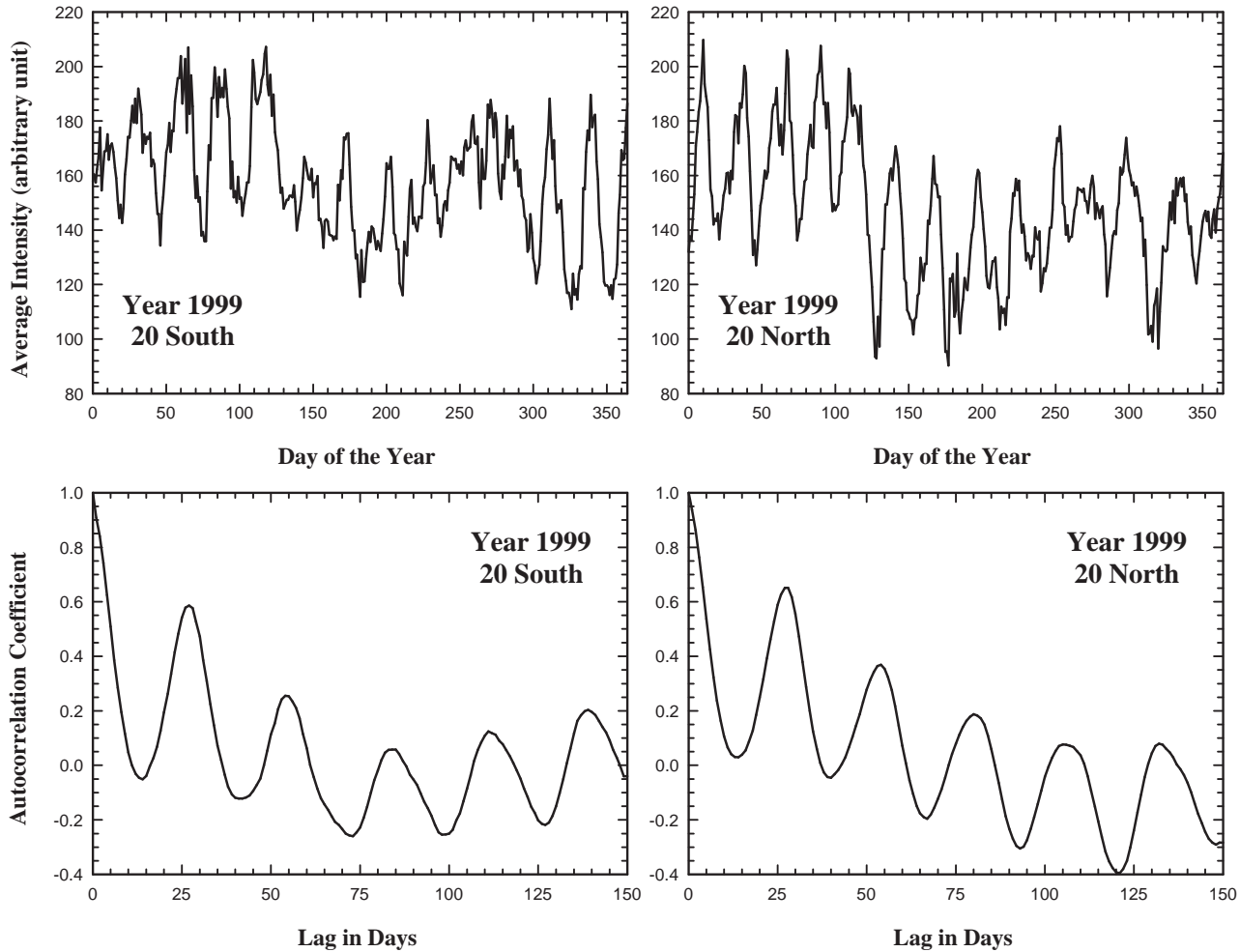}}
 \caption{The two panels on the top are typical examples of the time series of average soft X-ray flux of the year 1999 at latitudes 20\degr south and 20\degr north. The two panels on the bottom are their autocorrelograms. The peaks of these autocorrelogram are used in the present analysis to determine synodic rotation period.}
 \end{figure*}

There have been a series of studies on the coronal rotation rate employing Soft X-Ray (SXR) data \citep{b19,b14} in the past few decades, yet the picture is far from being completely clear. However, the extensive SXR database acquired by {\it Yohkoh} Soft X-ray Telescope (SXT) (extending almost the entire solar cycle) contains a wealth of information to be analyzed about the variation of coronal rotation rate with time and latitude. The rotation of the solar corona is still a matter of great interest because of its unique latitudinal and height profiles, temporal and spatial variations and its deviation from rigid rotation to the latitudinal-dependent differential rotation.

Observations of the solar coronal rotation period, in the last decades, have shown variations between rigid rotation to differential rotation profile across latitudes. \citet{b21} and \citet{b22} have used the data obtained by the {\it Yohkoh}/SXT to study the rotation rates of the corona and have reported that the soft X-ray corona does not show any significant differential rotation; it means the corona is more rigid than was thought before, compared to the photosphere or chromosphere.

Recently, by making use of X-ray bright points (XBPs) {\citet {b13} determined the coronal rotation rate employing solar full disc (SFD) images  obtained through the soft X-ray telescope (SXT) on the {\it Yohkoh} and the X-ray telescope (XRT) on the {\it Hinode} solar observatory. {\citet {b13} found that the corona does rotate differentially throughout the solar magnetic cycle, in the same way as its neighbouring lower atmospheric layers do, but it does not depend on the phases of the solar magnetic cycle.

After analyzing the synoptic photoelectric observations of the green (Fe XIV emission lines at 530.3 nm) and red (Fe X emission line at 637.4 nm) coronal lines for the periods 1973 to 2000 (Fe XIV) and 1984 to 2000 (Fe X), respectively, \citet{b17} and \citet{b1,b2} concluded that the rotation of corona shows more compactness than the rotation of the sunspots. The presence of a weak, solar cycle dependent differential rotation in both the spectral lines was also reported in their study. It indicates that the corona shows rigid rotation near the solar minimum and differential rotation near the solar maximum.

On the basis of the analysis of the coronal bright points observed on solar full-disc filtergrams at 19.5 nm (Fe XII line) wavelength with the extreme ultraviolet imaging telescope (EIT) on board the {\it SoHO} spacecraft, {\citet {b12}} found that the rotation of coronal bright points closely follows the latitudinal rotation profile of the photospheric magnetic field and, simultaneously, it was also shown that coronal features at different heights in the corona may exhibit different rotation rates. The SFD filtergrams from {\it SoHO} EIT (Fe XV) instrument at 28.4 nm were also used by {\citet {b6}} to analyze solar differential rotation by tracing coronal bright points. {\citet {b6}} determined through the procedure based on {\it IDL}, that the differential rotation profile corresponds roughly to the rotation of sunspot groups.

\begin{table*}
 \centering
 \begin{minipage}{112mm}
  \caption{Sidereal rotation periods of soft X-ray corona obtained through autocorrelation analysis for each latitude bins on SFD images for the years 1992-2001.}
  \begin{tabular}{@{}lcccccccccc@{}}
  \hline
   Latitude 	& \multicolumn{10}{c}{Years} \\
   bins at		&	1992	&	1993	&	1994	&	1995	& 1996  &  1997 &	1998	&	1999  &	2000	&	2001\\       
  \hline
	80\degr S		&	26.4	&	27.7	&	25.1	&	25.7	&	26.0	&	26.9	&	27.0	&	28.0	& 26.0	&	26.4\\
	70\degr S		&	26.4	&	27.1	&	25.1	&	25.6	&	26.0	& 26.0	& 27.0	& 27.7	& 25.4	& 26.4\\
	60\degr S		&	26.5	&	26.5	&	25.7	&	25.6	& 25.1	& 26.0	& 26.9	& 27.2	& 25.1	& 25.1\\
  50\degr S		&	25.7	&	27.2	&	25.1	&	25.6	&	25.6	& 26.0	& 26.7	& 26.0	& 25.1	& 24.9\\
  40\degr S		&	25.6	&	25.1	&	25.1	&	25.7	&	25.1	& 26.0	& 26.5	& 26.2	& 24.3	& 25.1\\
  30\degr S		&	26.0	&	25.1	&	25.3	&	25.1	&	24.6	& 26.0	& 26.0	& 25.8	& 24.3	& 25.6\\
  20\degr S		&	25.6	&	24.6	&	25.4	&	24.7	& 24.8	& 25.1	& 25.8	& 25.8	& 24.7	& 25.6\\
  10\degr S		&	24.9	&	25.1	&	25.1	&	24.7	&	24.8	& 25.4	& 25.7	& 25.7	& 24.0	& 24.3\\
  Equator			&	24.7	&	25.7	&	24.8	&	25.1	&	24.6	& 24.3	& 25.0	& 25.3	& 24.3	& 24.3\\
  10\degr N		&	25.8	&	25.6	&	24.8	&	25.0	&	25.1	& 25.1	& 25.1	& 25.1	& 24.6	& 23.8\\
  20\degr N		&	25.7	&	25.1	&	24.8	&	24.8	&	24.6	& 25.1	& 25.1	& 24.6	& 24.9	& 24.3\\
  30\degr N		&	26.0	&	25.1	&	25.0	&	25.1	&	24.6	& 25.6	& 26.0	& 25.1	& 24.9	& 25.4\\
  40\degr N		&	25.8	&	25.6	&	25.1	&	25.1	&	24.6	& 26.0	& 26.0	& 26.4	& 25.3	& 25.1\\
  50\degr N		&	25.6	&	26.0	&	25.3	&	26.0	&	25.4	& 26.0	& 26.0	& 26.4	& 25.5	& 24.6\\
  60\degr N		&	25.1	&	26.3	&	25.7	&	24.9	&	24.7	& 26.4	& 27.3	& 27.3	& 25.5	& 25.3\\
  70\degr N		&	26.4	&	25.1	&	25.7	&	25.1	&	26.0	& 26.9	& 27.1	& 28.2	& 25.7	& 26.6\\
  80\degr N		&	26.4	&	26.0	&	25.6	&	25.7	&	25.6	& 26.9	& 26.6	& 28.2	& 25.4	& 26.9\\
  \hline
\end{tabular}
\end{minipage}
\end{table*}

Disk integrated solar radio fluxes from 1997 to 1999 at different frequencies are used for the study of coronal rotation at different heights \citep {b24, b20}. \citet{b20} have shown that coronal rotation depends on the heights in the corona. Using Nobeyama radio images at 17 GHz for the years 1999 to 2001, \citet {b7} demonstrated that the rotation profile of the lower corona across the latitude is somewhat shallower than the rotation profile of layers below it.

From a detailed study of the data from {\it LASCO}, \citet {b15} concluded that the rotation of the corona displays a radially rigid rotation of 25.57 days sidereal period from 2.5 $R_{\sun}$  to $>$ 15 $R_{\sun}$.

The solar corona on the full disc can be best seen in soft X-ray images. This is because, in X-rays radiations, the corona is optically thin but, since the photosphere is comparatively less hot than the corona and therefore does not emit shorter wavelengths, soft X-rays radiations from the solar corona's full disc appear as bright emissions against a dark background of photosphere. Such high contrast images obtained in soft X-rays make it easier to track the coronal features present anywhere on the solar disc and to determine spatial and temporal variations by using the time series SXR data.

The paper is structured as follows. In section 2, we discuss the source and span of the data which we have used in this study. In its subsection, we briefly give the technique used to obtain the time series from the pre-processed solar full disc images. In section 3, we present the method used to determine the rotation period at various latitudes from such time series. In section 4, the rotation rate and its differential profile is compared with recent published results. It also includes the discussion on the dependency of the rotation rate on the activity cycle. The paper ends with conclusions.

\section{The Observations}

The {\it Yohkoh}, which was the Japanese solar observatory satellite, used many telescopes and other detectors to observe the X-rays that are produced by the Sun \citep{b27}. The soft X-ray telescope (SXT) of 1.54 m focal length which forms X-ray images on a $1024 \times 1024$ virtual phase CCD detector was one of them. This telescope used grazing incidence mirrors (Modified Wolter type I) to form SXR images of the Sun in the wavelength range 0.3-4.5 nm (selectable with filters) on a CCD sensor.  The SXT produces full and partial solar disc images with an angular resolution of 5 arc-sec and a time resolution of 2 sec (minimum) \citep {b26}. The web-accessible database, called SFD images, is primarily in the form of digital whole-Sun images in $512 \times 512$ pixel sizes, available at a cadence of one image per day, for dates spanning the entire {\it Yohkoh} mission (since Oct 1991 until Dec 2001), with very few data gaps ($<6\%$).

\subsection{Data Reduction}

The SXT is a broadband instrument (0.3-4.5 nm) in SXR, which is designed for observing and identifying active regions, diffuse and/or unresolved non-active regions and coronal holes (i.e., tracers of the SXR signal), explicitly. The studies \citep{b8} on coronal structure indicate that large-scale features can persist for several synodic rotations. The persistence of such tracers justifies the use of the autocorrelation analysis to determine the characteristic rotation rates \citep{b9}. The technique of auto correlation is not new in studies of the rotation period. It has often been used in previous studies for determining the coronal rotation by \citet {b17,b25} and \citet{b21}. The autocorrelation technique has proved to be useful in identifying any prominent signals present in the harmonic content of the data.

The basic assumption used in this paper is that the SXR flux modulation is evident from the periodic passage of features which are persistent for more than one synodic rotation. \citet {b8} and \citet{b17} have shown (using autocorrelation analysis) that tracer lifetimes can be longer than one solar rotation, at a minimum, and can be longer than two or more solar rotations in some cases. Here preprocessed images of SXT (Yohkoh) for the period 1991 - 2001  are used. This image processing (private communication, L. Acton 2010) involves,

\begin{enumerate}
	\item 
subtraction of CCD dark signal from images; 
	\item
logarithmical compression to 8 bit (256 intensity level) images after one or two bits have been truncated from the bottom to suppress noise in the faintest parts; 
	\item
if there is a very bright flare there is an upper limit threshhold and all pixels above this are set to 256; and
	\item
to keep the solar minimum images from being too dim there is a solar cycle correction applied to increase the overall brightness of images obtained when the Sun is very quiet.
\end{enumerate}

In the ten years data set we discard corrupt images. The number of corrupt images in the data set is very small and it varies from 0 to 19 per year. Thus the data set is reasonably continuous and the data has been recorded at regular intervals. Moreover, we use linear interpolation to replace the discarded images. The modified data sets are divided into one-year segments which start with the beginning of each calendar year. One year length is chosen because it encompasses several cycles of solar rotation. To produce a time-series from each one-year dataset, we utilize full disk SXR images, of $512 \times 512$ pixel sizes, per day. Then, seventeen latitude bins are selected at intervals of 10\degr on the heliocentric latitude lines in each image, from 80\degr N to 80\degr S. The width of each bin is just two pixels and the length is such as to include entire pixels of the image at each latitude. A sufficient number of pixels in each bin are necessary so that statistically credible estimates of the mean SXR flux can be made. The dimensions of each bin are selected, so that it can incorporate all the coronal features which persist for more than one synodic rotation at any latitude, and which track down to estimate SXR flux modulation. 

\begin{figure*}
  \centering{
  \includegraphics[width=\textwidth]{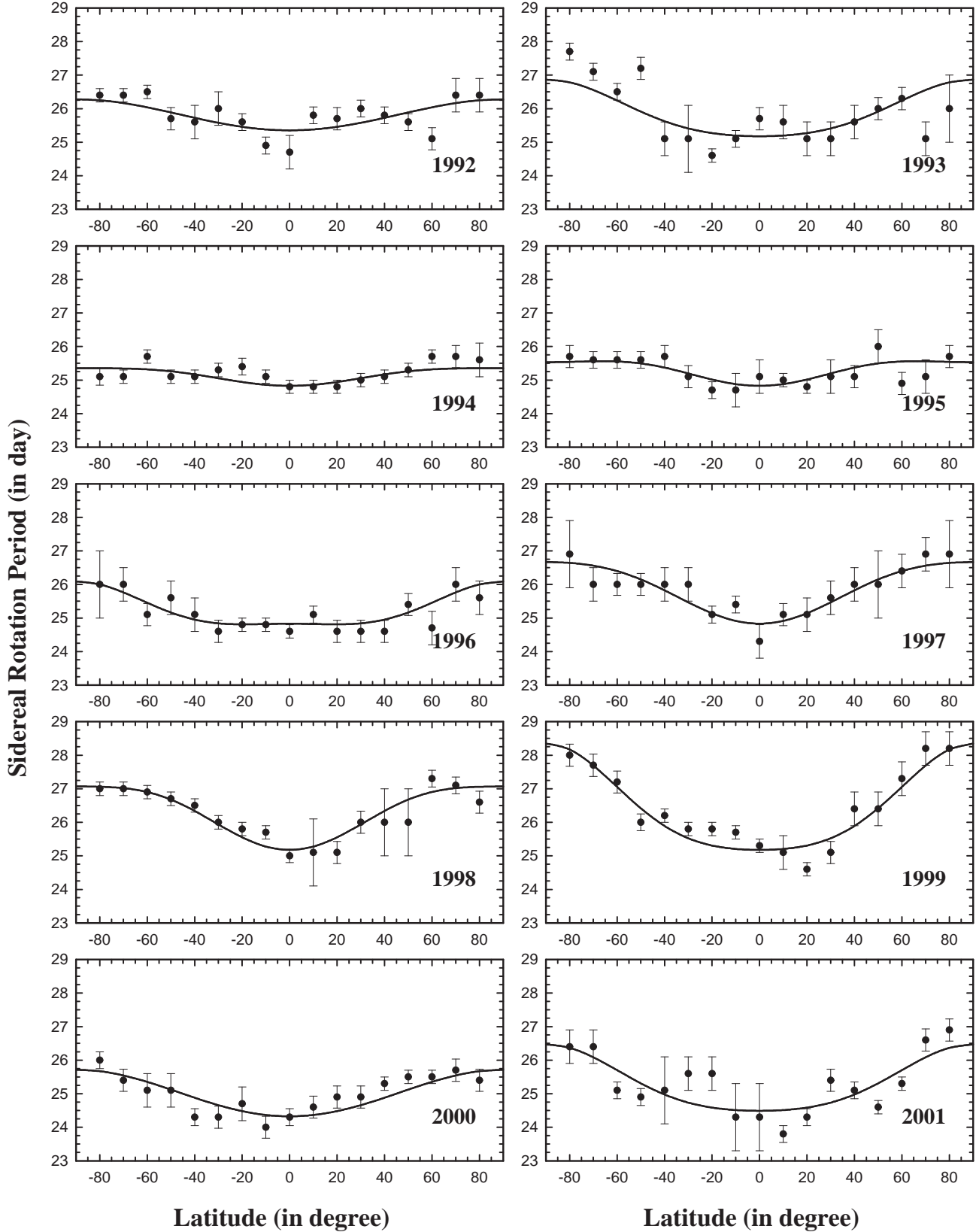}}
 \caption{Sidereal rotation period as a function of latitude obtained from the SFD images of Yohkoh SXT of the years 1992-2001. The size of error bars depends on the selection of the peak in determination of the rotation period. The least square curve fitted in the derived profile of the sidereal rotation period are also shown.}
 \end{figure*}

Each time series contains information from the specified bin area represented by the selected pixels. In this way, the averaged SXR flux from each latitude bin is converted into a time-series. The time-series so produced, are analyzed for each latitude bin, giving a total of 17 time-series datasets. We obtained the autocorrelation coefficient of the average pixel intensity modulation for each bin by utilizing standard subroutines used in IDL software. The autocorrelogram so obtained is utilized to determine the synodic rotation period of the corona as a function of latitude.

It is important to mention here that the latitude bin at 90\degr north or south could not be framed by the bin selection method used in this work, as it would have point dimensions. Statistically reliable information could not  have been retrieved from such latitude bins.

\begin{table}
 \centering
 \begin{minipage}{84mm}
  \caption{The coefficients (in the equation for the sidereal rotation rate) $A$, $B$ \& $C$ and their standard error $E_A$, $E_B$ \& $E_C$ in deg/day, for the years 1992-2000 from Yohkoh SXT data. The \textit{rms} deviation between the derived data and curve fitted data and the annual sunspots number (SSN) are also listed to compare with the coefficients.}
  \begin{tabular}{@{}lccccc@{}}
\hline
Year 				& \multicolumn{3}{c}{Coefficient} 								&	\textit{rms}			&	SSN\\
     				& $A \pm E_A $    &  $B \pm E_B  $ 		&	$C \pm E_C  $			&	dev.			&	\\       

\hline
1992				&	14.2 $\pm$ 0.1	&	$-$0.5 $\pm$ 0.7	& $-$0.0 $\pm$ 0.7  & 0.4 			& 94.5\\
1993				&	14.3 $\pm$ 0.2	&	$-$0.4 $\pm$ 1.0	& $-$0.5 $\pm$ 1.0  & 0.6 			& 54.7\\
1994				&	14.5 $\pm$ 0.1	&	$-$0.6 $\pm$ 0.4	& $+$0.3 $\pm$ 0.4  & 0.2 			& 29.9\\	
1995				&	14.5 $\pm$ 0.1	&	$-$1.0 $\pm$ 0.5	& $+$0.6 $\pm$ 0.5  & 0.3 			& 17.5\\	
1996				&	14.5 $\pm$ 0.1	&	$+$0.2 $\pm$ 0.6	& $-$0.9 $\pm$ 0.6  & 0.3 			& 8.6\\
1997				&	14.5 $\pm$ 0.1	&	$-$1.6 $\pm$ 0.5	& $+$0.6 $\pm$ 0.5  & 0.3 			& 21.5\\
1998				& 14.3 $\pm$ 0.1	&	$-$1.9 $\pm$ 0.5	& $+$0.9 $\pm$ 0.5  & 0.3 			& 64.3\\
1999				&	14.3 $\pm$ 0.1	&	$-$0.5 $\pm$ 0.6	&	$-$1.1 $\pm$ 0.6  & 0.3 			& 93.3\\
2000				&	14.8 $\pm$ 0.1	&	$-$0.8 $\pm$ 0.6	&	$+$0.0 $\pm$ 0.6  & 0.3 			& 119.6\\
2001				&	14.7 $\pm$ 0.2	&	$-$0.4 $\pm$ 1.0	&	$-$0.7 $\pm$ 1.0  & 0.5 			& 111.0\\
\textbf{Avg}& 14.5 $\pm$ 0.1  & $-$0.8 $\pm$ 0.6	& $-$0.1 $\pm$ 0.6  & 0.4 			&    \\
\hline
\end{tabular}
\end{minipage}
\end{table}

\section{Data Analysis \& Results }

The temporal plots of the time series obtained from the daily X-ray images for each of the latitude bins have a variation which shows that rotational modulation exists. One typical example of the year 1999 for 20\degr N and 20\degr S is shown in Figure 1 (top panels). The systematic rotational modulation due to the solar rotation can be estimated in all the data sets from their autocorrelogram. Figure~1 (bottom panels) is a typical example of such an autocorrelogram obtained at 20\degr south and 20\degr north, for the year 1999. These kinds of time series are used to derive the autocorrelation coefficient. Most of the curves of the autocorrelogram obtained in this way at different latitudes, having different bin sizes, usually demonstrate a smooth nature with a fair amount of correlation. This may be due to the persistence of coronal tracers for the long term on the solar disc at that latitude. The autocorrelogram obtained at some of the latitudes shows such a good correlation that, peak after peak, the modulation remains uniform and even the fifth peak is as clear as the first one (as shown in the bottom panels of Figure~1). In some of the plots, the first few peaks show uniformity but this uniformity is soon lost with the upcoming peaks. It seems that short term, less prominent tracers and higher latitude locations (because the bin size becomes progressively smaller) affect adversely the smoothness of the curves. The number of cyclic variations as well as the magnitude of the autocorrelation coefficient also decreases. In spite of these, the first peak of the autocorrelograms seems to be clear and is useful to estimate the synodic rotation at a particular latitude. The synodic rotation period can be obtained by determining the position of the peaks on the lag (in days) axis. If the position of the first peak is chosen to determine the synodic period then there will be an error of one day. But this error can be reduced by selecting the position of the farthest peak. If we chose the position of the second, third, fourth or fifth peak, then the error will be reduced to 1/2, 1/3, 1/4 or 1/5 day, respectively. Hence, to determine the synodic rotation period with the maximum possible accuracy, we have to choose the position of a peak as far as possible. But the curve should show smoothness and cyclic nature up to this peak. Therefore, the accuracy varies from one latitude bin to another. We followed this approach to determine the synodic period and then the sidereal rotation period is calculated for each year from 1992 to 2001. The sidereal rotation periods so obtained are listed in Table~1 for comparison.

Most of the measurements of solar rotation as a function of latitude, traditionally fit a functional curve by least square method. This functional form has been in the even powers of sine of the latitude, as given below

\begin{equation}
\Omega(\psi)=A + B \sin^2\psi + C \sin^4\psi
\end{equation}

where $\Omega(\psi)$ is the solar rotation rate at any solar latitude $\psi$, the parameter $A$ represents the equatorial rotation rate, $B$ and $C$ measure the differential rate with $B$ representing mostly lower latitudes and $C$ representing mainly higher latitudes {\citep {b8, b4}}. These are listed in Table~2. 

The coefficients $B$ and $C$ are coupled by an inverse correlation, which is termed as cross talk in the coefficients and makes comparison more difficult between coefficients obtained from different measurements. The detailed mathematical procedures to remove and/or reduce cross talk are described in a series of papers e.g. {\citet{b18}} and {\citet{b23}} etc.. In some papers e.g. {\citet{b10}} and {\citet{b3}}, $C=0$ and in others e.g. {\citet{b6}} and {\citet{b13}} $C=B$ is stated to be the solution for cross talk removal; however, these arguments have no justification. In view of this, in the present work we do follow the tradition of fitting the curve using equation~1 and obtain coefficients $A$, $B$ and $C$ and also make a table for the same as these are used to derive the profiles. It may be reasonable to consider the temporal variation of any coefficient separately, but for the inter comparison of our results with others we will use the full latitudinal profiles obtained by various observations. In the present work the dataset is available from the latitude bins ranging from the equator up to 80 degrees, of both hemispheres.

The rotation rate $\Omega$, at any latitude $\psi$ could be converted into rotation period $T$ simply by

\begin{equation}
T(\psi)=\frac{360\degr}{\Omega(\psi)}
\end{equation}

where $T(\psi)$ is in day and $\Omega$ in degree/day.

\begin{figure*}
\centering{
  \includegraphics[width=\textwidth]{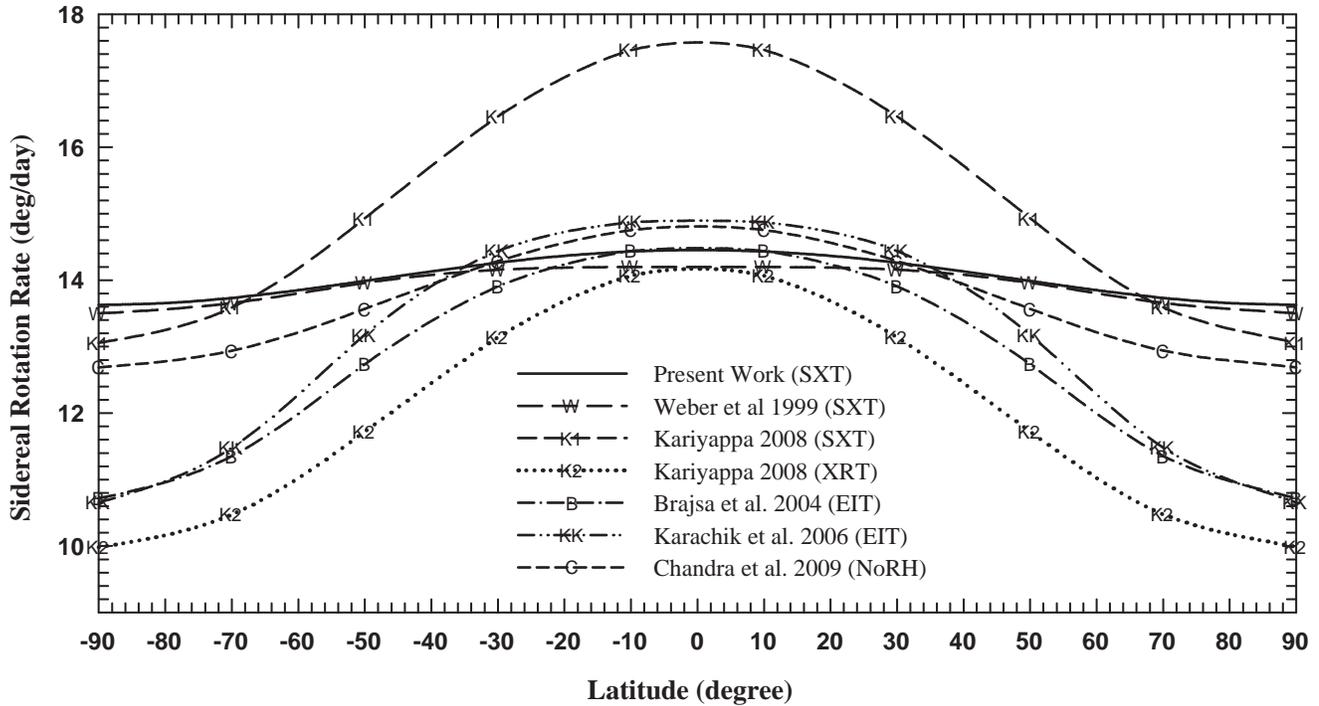}}
 \caption{Comparison of sidereal differential rotation profiles obtain using coefficient $A$, $B$ and $C$ of various data sources.}
 \end{figure*}

\begin{table*}
 \centering
 \begin{minipage}{135mm}
 \caption{Sidereal rotation parameters $A$, $B$ and $C$ (in $deg/day$), obtained in present work with ({\it Yohkoh}/SXT) along with others by different observations. These values are used to derive the curves of Figure~3. All the values are given only up to first place of decimal.}
 \begin{tabular}{@{}lllllll@{}}
 \hline
  Data  		&		  			& \multicolumn{3}{c}{Coefficients} 	& References 	&Period \\
  Source    &						& $A$   & $B$   &$C$ 	&            	&\\
 \hline
  Yohkoh/SXT&0.3-4.5 nm	&14.5		&-0.8		&-0.1	&present work	&1992-2001 \\
  Yohkoh/SXT&0.3-4.5 nm &14.2		&-0.0		&-0.7	&\citet{b21} 	&1992-1997 \\
  Yohkoh/SXT&0.3-4.5 nm	&17.6		&-4.5		&0.0	&\citet{b13} 	&1992-2001 \\
  Hinode/XRT&0.2-20 nm	&14.2		&-4.2		&0.0	&\citet{b13} 	&Jan, Mar \& Apr 2007 \\
  SOHO/EIT	&28.4 nm  	&14.5		&-1.9		&-1.9	&\citet{b6}	 	&Jun 1998-May 1999 \\
  SOHO/EIT	&19.5 nm  	&14.9		&-1.1		&-3.2	&\citet{b12}	&Apr \& Jul 1996, May 2005\\
  NoRH			&1.76 cm  	&14.8		&-2.1		&0.0	&\citet{b7}  	&1999-2001 \\

 \hline
\end{tabular}
\end{minipage}
\end{table*}

In Figure~2, the sidereal rotation periods with respect to latitudes are plotted in ten panels for the years 1992 to 2001, separately. The error bars associated with the sidereal rotation periods show the accuracy in its determination. It is evident from Figure~2, in spite of significant scatter, that the corona in SXR exhibits a certain degree of differential rotation. The latitude differential is maximum in 1999 and minimum in 1994. These profiles have error bars ranging from $\pm $ 0.2 to $\pm$ 1 day. Least square curve is fitted to the observed profile using equation ~1. We obtain the corresponding coefficients $A$, $B$ and $C$ (with their standard errors) of coronal rotation rate of each year. These coefficients and their standard errors are listed in Table~2. The \textit{rms} deviation between the observed and fitted profile of the sidereal rotation period is also calculated and listed in Table~2. The small \textit{rms} deviation (in the range of 0.2 to 0.6) indicates that the fitting is reasonably good.

\section{Discussion}

\subsection{Comparison with other data}

\begin{figure*}
\centering{
  \includegraphics[width=\textwidth]{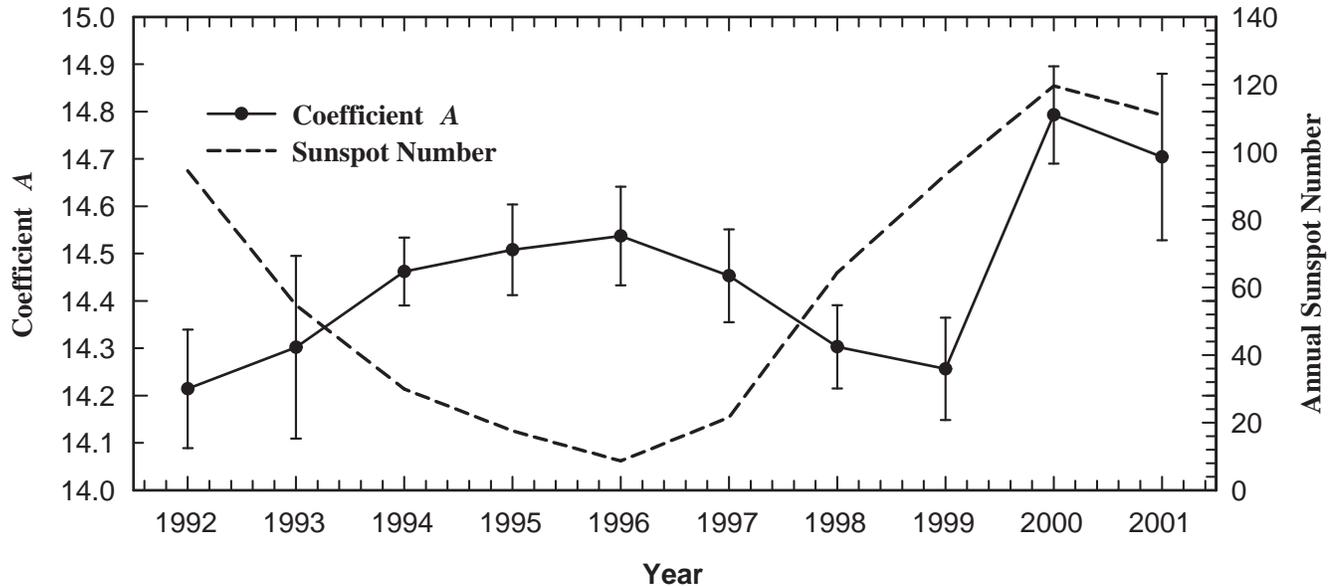}}
 \caption{Comparison of coefficient $A$ with the yearly mean of sunspot numbers.}
 \end{figure*}

In recent years, various methods have been employed to estimate the solar rotation profile using satellite SFD images. The SFD images obtained through {\it Yohkoh} SXT \citep {b21,b13}, {\it Hinode} XRT \citep{b13} or {\it SoHO} EIT, etc., were extensively used for this purpose. Many ground based observatories are also observing full disk images at different frequencies on a regular basis. The Nobeyama radioheliogram at 17 GHz, obtained from such an observatory, is also being used for estimating coronal rotation \citep{b7}. It is therefore, natural to compare the present work with these recently published results. 

The coefficient $A$ is the measure of equatorial rotational velocity, whereas coefficients $B$ and $C$ are the estimates of differential gradients of coronal rotation at lower and higher latitudes, respectively. We list these coefficients along with different published results in Table~3. The curves of latitude profiles of sidereal rotation rates are also obtained using these coefficients. The comparisons between the curves are shown in Figure~3. The data used by these are as follows;
\begin{enumerate}
	\item SXR flux modulation using time-series latitude bins of {\it Yohkoh}/SXT images for the period 1992-2001, in present work (----------------),
  \item Harmonic filtering using Lomb-Scargle periodogram of {\it Yohkoh}/SXT images by \citet {b21}, for the period 1992-1997 (--- W --- W ---),
  \item X-ray bright points (XBPs) of {\it Yohkoh}/SXT SFDs data for the period of 1992 to 2001, obtained recently by \citet {b13} (-- K1 -- K1 --),
  \item X-ray bright points (XBPs) of {\it Hinode}/XRT SFDs data for the three months in 2007, obtained again by \citet {b13} (.. K2 .... K2 ..),
  \item Coronal bright points (CBPs) of {\it SoHO}/EIT SFDs data for the period from Jun 1998 to May 1999 derived by \citet {b6} (--B-- . --B--),
  \item Coronal bright points (CBPs) of {\it SoHO}/EIT SFDs data of Apr 16, July 28 of 1996, May 05-06, 2005 derived by \citet {b12} (--KK-- .. --KK--), and
  \item Time series autocorrelation analysis on latitude bins of radio images at 17 GHz obtained from Nobeyama Radioheliograph for time interval from 1999 to 2001 \citep {b7} (- - C - - C - -).
\end{enumerate}

\subsection{Coronal equatorial rotation rate}

As can be seen from Figure~3, the mean rotation rate at the equator (derived through the coefficient $A$ of all the years) matches well, in the error limit, with most of the observations discussed before (except with \citet{b13}). Interestingly, the equatorial rotation rate is found to be almost the same, when compared with the extreme ultraviolet (EUV) chromospheric datasets \citep {b6,b12} and coronal radio data at 17 GHz \citep {b7}. But, it is more than the SXT rotation rate obtained by using the harmonic filtering method \citep {b21} and the XRT rotation rate derived from XBPs analysis \citep {b13}. The mean rotation rate suggests that near the equator the SXR corona rotates almost at a speed close to the speed of the photosphere \citep{b10,b3} and chromosphere \citep{b6}.

The curve of XRT dataset, due to {\citet {b13}}, does not match at almost all the latitudes, except at the equator with the curve of {\citet {b21}}, with any of the cases included in this paper. Whereas, the fit of SXT data set matches only at the higher latitudes with both results obtained through SXT data sets (present work \& \citet{b21}) and NoRH radio data. This may be due to the cross talk between the coefficients when they are mutually compared with different measurements, since only the first two coefficients were used in the polynomial with XRT and SXT dataset.

\subsection{Coronal differential rotation profiles}

The curves due to \citet {b6} and \citet {b12}, at EUV wavelengths, show large differential gradient values, at middle and higher latitudes. But they, themselves, closely match with each other (in spite of choosing different remedies to avoid cross talk between coefficients $B$ and $C$). With radio images at 17 GHz \citep {b7}, the differential gradient profile seems to be quite high up to the middle latitudes, and low at higher latitudes. Since \citet {b7} have not used data of higher latitudes in both hemispheres and have taken only the first two terms of the polynomial expansion, hence, the curve due to \citet {b7} may be reliable only up to the middle latitudes. This may lead to higher variance at higher latitudes. The profiles seem to indicate the fact that the corona displays a variety from almost rigid \citep{b15} to reasonably differential rotation. The curve due to {\citet {b21}} is found closest to the present study at the middle and higher latitudes.

\subsection{Variation in coronal equatorial rotation as a function of the phase of the solar cycle}

The coefficient $A$ represents the coronal equatorial rotation velocity and this seems to depend on the phases of the solar cycle (see Table~2). Figure~4 shows the temporal variation of coefficient $A$ (continuous line with error bars) derived from the present work and the annual mean sunspot number (dotted line). The error of $A$ varies from $\pm$ 0.1 to $\pm$ 0.2. The temporal variation of $A$ is apparent. In the descending phase of the solar cycle 22 (1992-96), coefficient $A$ increases gradually. In the ascending phase of cycle 23, {\it i.e.}, in the years 1996 to 1999, coefficient $A$ decreases steadily. The rotation rate during the year 2000 is exceptionally high and is thus, out of phase with the rest of the year's data point. This was the year of solar maxima and hence some extraordinary activity might have enhanced the equatorial rotation velocity to such an extent. Similar enhancement in the year 2000 was also reported by \citet{b7}, from the analysis of the radio images at 17 GHz. The coefficients $B$ and $C$ do not show any systematic variation.

\section{Conclusions}

It is clear from Figure 3, that the equatorial sidereal rotation rate we obtained (14.5 deg/day as shown in Table 3) is similar to those obtained by the \citet{b6} (14.5 deg/day from Table 3) or by the \citet{b12} (14.9 deg/day from Table 3). Interestingly, they both used as tracers coronal bright points at chromospheric levels ({\it SoHO}/EIT). A similar inference can be drawn at the photospheric level (14.6 deg/day determined by \citet{b3} using sunspots as tracer). So from the present study, we infer that the soft X-ray corona at the equator rotates almost at the same rate as the lower atmospheric layers.

The differential profile as a function of latitude varies considerably throughout the period of the study, extending from 1992 to 2001. The differential gradient is found to be maximum in the year 1999 and minimum in 1994. The comparison shows that the curve of the present work using the {\it Yohkoh} dataset shows latitudinal dependent differential rotation but its mean gradient is less in comparision to most of the previously published results (except that of \citet{b21}). It implies that the rotation profile of soft X-ray corona across latitudes is shallower than that of chromosphere and photosphere. The result is important in the sense that SXR corona with {\it Yohkoh}/SXT dataset shows the same qualitative rotational behaviour as that obtained in some other studies, for example, \citet{b9, b8, b21}, employing different rotation measurement techniques.

The equatorial rotation rate in the solar corona shows a systematic variation with respect to the annual sunspots number. But the variation in its rate is anti-correlated with the sunspots number (except in the year 2000). This indicates the dependence of the coronal equatorial rotation velocity on the phases of the solar cycle.

\section*{Acknowledgments}

The authors wish to acknowledge the data full disc images in soft X-ray and annual sunspot numbers used in the present work. These were acquired from the web pages of {\it Yohkoh Solar Observatory\/} (YSO),{\it Yohkoh Data Archive Centre} (YDAC) and {\it National Geophysical Data Centre} (NGDC). The research at {\it Physical Research Laboratory} (PRL) is supported by the {\it Department of Space} (DOS), Government of India. The authors would like to express their sincere thanks to David E. McKenzie and Loren Acton of Montana State University for his valuable suggestions and discussion on the SXT observations.  The authors are grateful to the anonymous referee for his constructive comments and valuable suggestion for the improvement of the paper.


\bsp

\label{lastpage}

\end{document}